\documentclass[11pt,twoside]{article}

\usepackage{asp2006}
\usepackage{epsf}
\usepackage{psfig}
\usepackage{lscape}
\usepackage{natbib}
\begin{document}
\title{HII regions feeding the interstellar medium in M\,83}   
\author{Javier Blasco-Herrera, Kambiz Fathi, John Beckman, Leonel Guti\'errez, Andreas Lundgren, Joan Font, Olivier Hernandez, Claude Carignan}   
\affil{Stockholm Observatory, Sweden; IAC, Spain; UNAM, Mexico; ESO, Chile; LAE Montreal, Canada.}    



\section*{Overview}
In this study, we analyse the internal dynamics of star-forming {\sc H\,i\,i} regions and their efficiency in interacting with the ambient interstellar medium. We have selected the nearby (4.5 Mpc) nearly face-on spiral galaxy M\,83, and from {\sc GH$\alpha$FaS} \citep{Hernandezetal2008} Fabry-Perot observations of the central $3.4^{\prime}\times3.4^{\prime}$ region, extracted the integrated H$\alpha$ emission-line profiles of a sample of 136 isolated {\sc H\,i\,i} regions. We investigate the effect of the instrumental response and find that, although commonly this effect is subtracted quadratically \citep[e.g.,][]{Rozasetal1998}, the contribution of the instrument assuming Gaussian instrumental response is not adequate for deriving reliable emission line profiles. Instead, we convolve one Gaussian with the observed instrumental response \emph{before} fitting the result to the line profile. Quadratic subtraction overestimates the H$\alpha$ emission-line velocity dispersion by $\approx 6 \mbox{ km s}^{-1}$ (left panel in the figure). When performing the same study fitting two Gaussians, results show that the secondary component tends ($>80\%$ of the cases) to be  smaller and narrower, while the uncertainties in the determination of the parameters of \emph{both} Gaussians increase  by as much as a factor of three.

\begin{figure}[!h]
 \plottwo{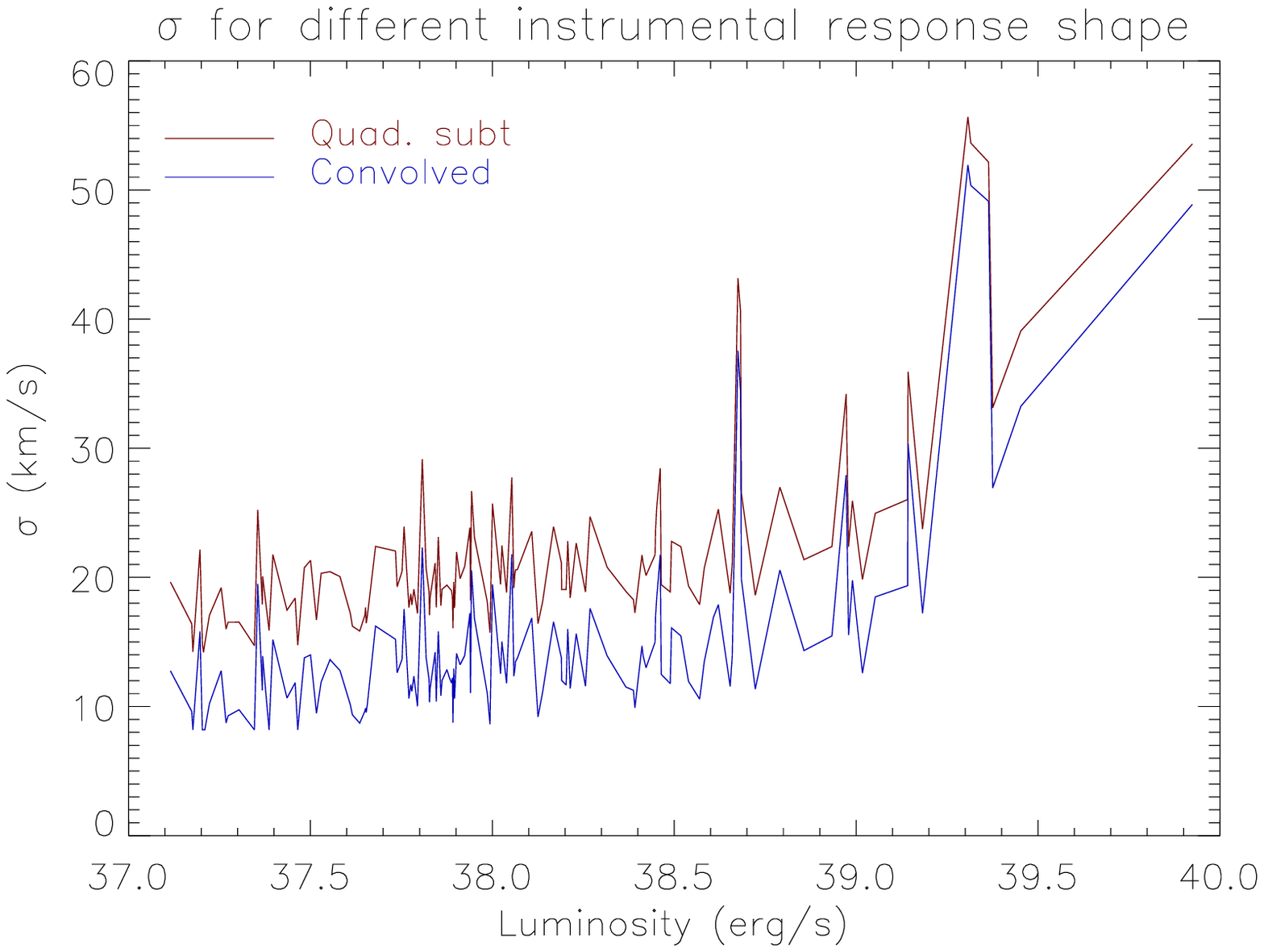}{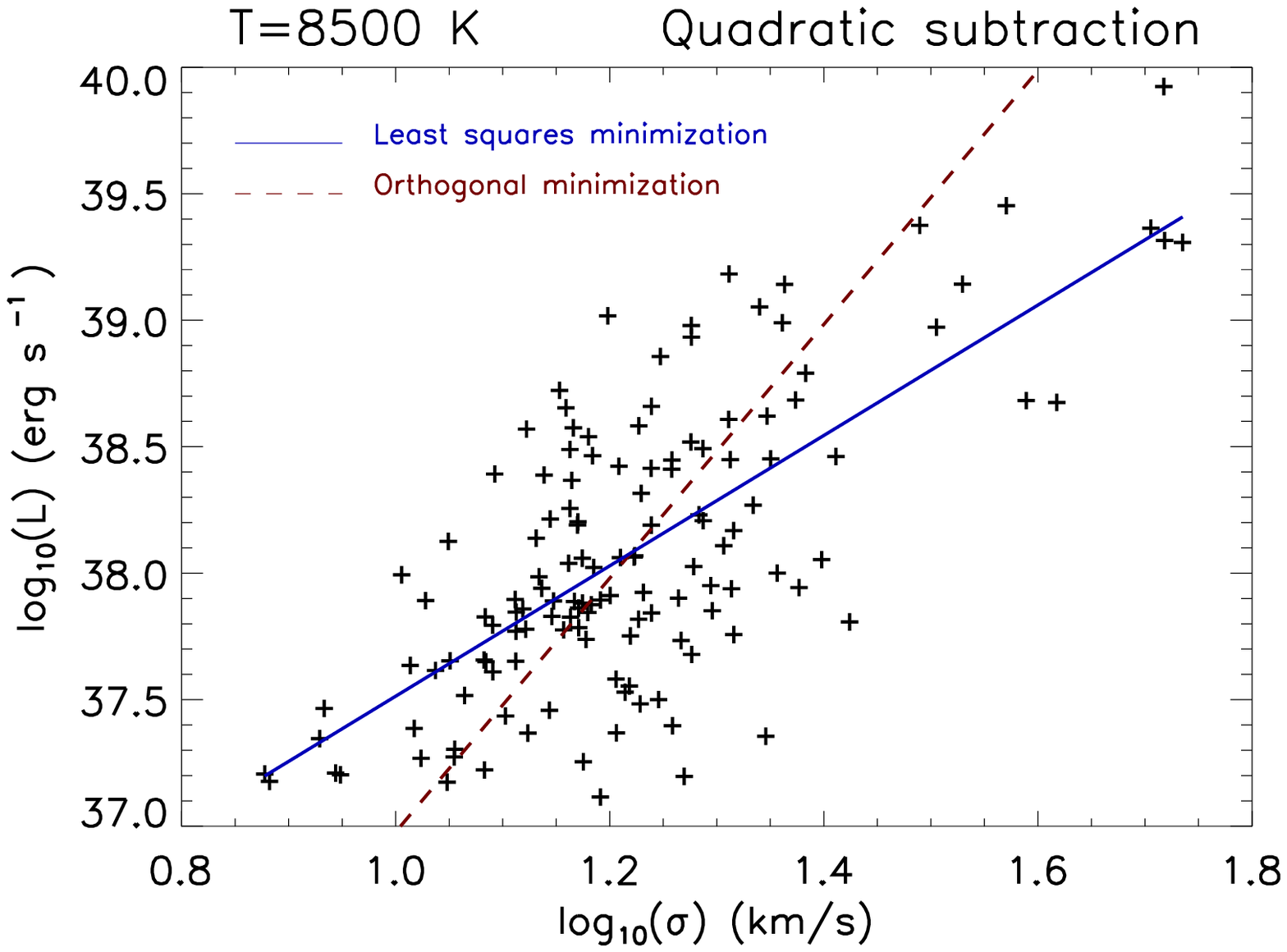}
\end{figure}

We investigate the relation between \emph{Luminosity} and H$\alpha$ \emph{Velocity dispersion} $\sigma$ (right panel), and reproduce the correlation found in previous studies \citep{Relanoetal2005,Rozasetal1998}, however, due to the large scatter in the derived $\sigma$ values, we find that the slope of the linear fit could vary between 2 and 5 depending on the minimisation strategy and assumptions about systematic errors caused by assumptions about the mean temperature and metallicity in {\sc H\,i\,i} regions. Our goal is to study these effects and to quantify the energetic output from {\sc H\,i\,i} regions, as well as estimate their age and evolution. A comprehensive analysis will appear in \citep{Blascoetal2010}.






\end{document}